\documentclass{article}

\usepackage{microtype}
\usepackage{graphicx}
\usepackage{subfigure}
\usepackage{booktabs} 
\usepackage{braket}
\usepackage{amssymb} 
\usepackage{bm}      
\usepackage{colortbl}   
\usepackage{xcolor}     
\usepackage{array}      

\definecolor{best}{RGB}{255,180,180}      
\definecolor{second}{RGB}{255,210,170}    
\definecolor{third}{RGB}{255,245,180}     

\def\our{QuantumGS}

\usepackage{hyperref}

\usepackage[preprint]{icml2026}
\usepackage[utf8]{inputenc}
\usepackage[T1]{fontenc}

\usepackage{amsmath}
\usepackage{amssymb}
\usepackage{mathtools}
\usepackage{amsthm}

\def\our{QuantumGS}

\usepackage[capitalize,noabbrev]{cleveref}

\theoremstyle{plain}

\theoremstyle{definition}

\theoremstyle{remark}

\usepackage[textsize=tiny]{todonotes}

\icmltitlerunning{\our{}}

\begin{document}

\twocolumn[
\icmltitle{\our{}: Quantum Encoding Framework for Gaussian Splatting}



\icmlsetsymbol{equal}{*}

\begin{icmlauthorlist}
   \icmlauthor{Grzegorz Wilczy\'nski}{yyy,sch}
    \icmlauthor{Rafa\l{} Tobiasz}{yyy,sch}
    \icmlauthor{Pawe\l{} Gora}{yyy}
    \icmlauthor{Marcin Mazur}{yyy}
    \icmlauthor{Przemys\l{}aw Spurek}{yyy,sch}
\end{icmlauthorlist}

  \icmlaffiliation{yyy}{Jagiellonian University}
  \icmlaffiliation{sch}{IDEAS Research Institute}

\icmlcorrespondingauthor{}{przemyslaw.spurek@uj.edu.pl}

\icmlkeywords{Machine Learning, ICML}

\vskip 0.3in
]



\printAffiliationsAndNotice{\icmlEqualContribution} 

\begin{abstract}
Recent advances in neural rendering, particularly 3D Gaussian Splatting (3DGS), have enabled real-time rendering of complex scenes. However, standard 3DGS relies on spherical harmonics, which often struggle to accurately capture high-frequency view-dependent effects such as sharp reflections and transparency. While hybrid approaches like Viewing Direction Gaussian Splatting (VDGS) mitigate this limitation using classical Multi-Layer Perceptrons (MLPs), they remain limited by the expressivity of classical networks in low-parameter regimes. In this paper, we introduce \our{}, a novel hybrid framework that integrates Variational Quantum Circuits (VQC) into the Gaussian Splatting pipeline. We propose a unique encoding strategy that maps the viewing direction directly onto the Bloch sphere, leveraging the natural geometry of qubits to represent 3D directional data. By replacing classical color-modulating networks with quantum circuits generated via a hypernetwork or conditioning mechanism, we achieve higher expressivity and better generalization. Source code is available in the supplementary material. Code is available at \url{https://github.com/gwilczynski95/QuantumGS}
\end{abstract}

\begin{figure}[t]
\centering
\includegraphics[width=\linewidth]{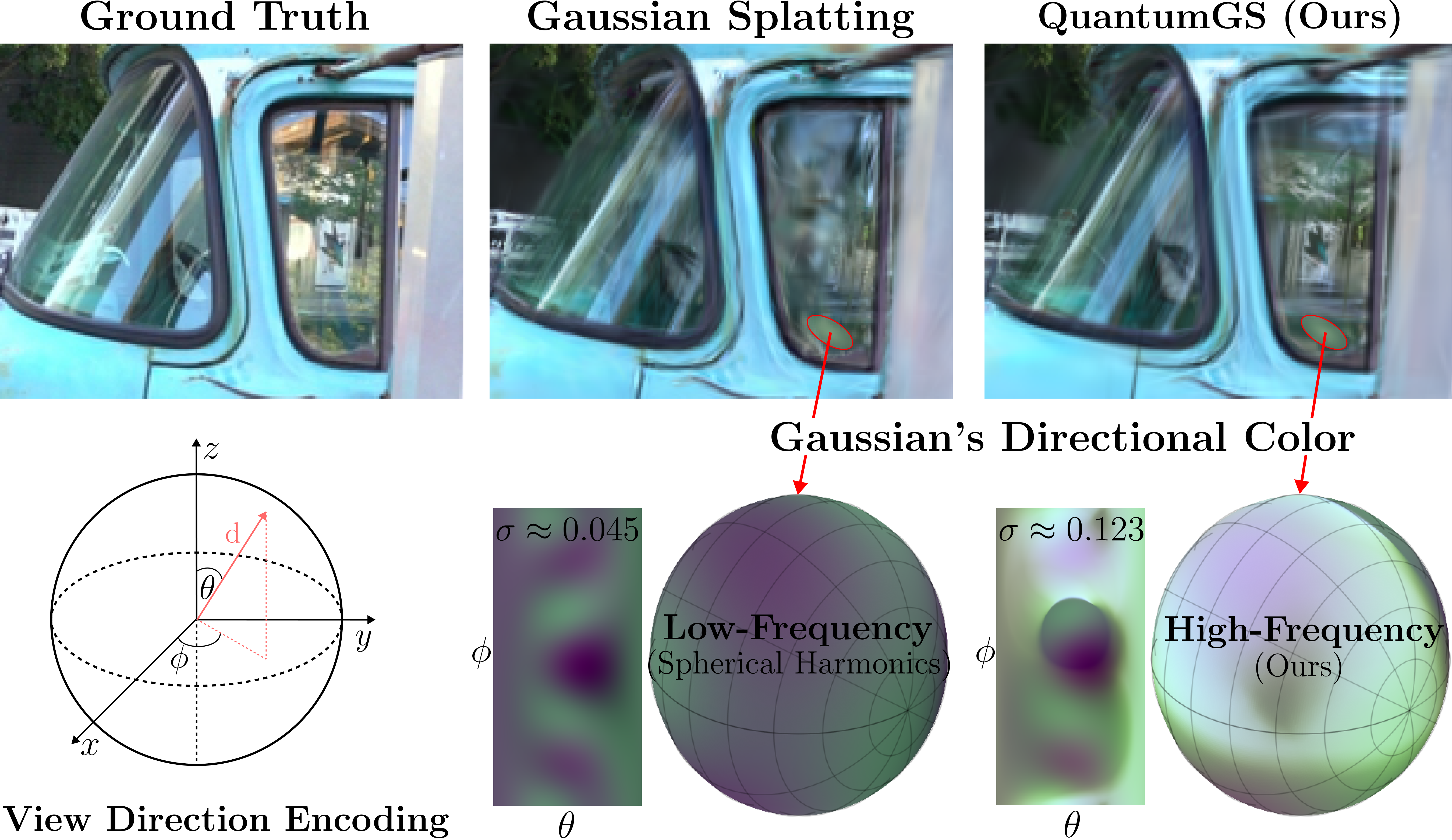}
\caption{
{\em Top}: Truck scene from Tanks and Temples~\cite{knapitsch2017tanks} demonstrates complex transparency. Standard 3DGS blurs the poster behind the windshield due to low-frequency spherical harmonics. \our{} preserves high-frequency view-dependence, recovering background visibility. {\em Bottom}: Directional color response of a single Gaussian. Unlike smooth SH patterns (middle), Bloch-sphere encoding (right) learns complex, irregular responses (e.g., central dark lobe), enabling precise light transmission modeling.
}
\label{fig:teaser}
\end{figure}

\section{Introduction}
\label{sec:introduction}

Recent advances in neural rendering have shifted from implicit coordinate-based representations like Neural Radiance Fields (NeRF)~\cite{vanilla_nerf} toward explicit, point-based methods. Leading this evolution, 3D Gaussian Splatting (3DGS)~\cite{vanilla_3dgs} achieves real-time rendering of complex scenes by modeling geometry as anisotropic 3D Gaussians. Despite its excellent speed-quality trade-off, standard 3DGS struggles with high-frequency view-dependent effects like sharp specular highlights, glossy surfaces, and transparency variations due to reliance on low-order spherical harmonics for color modulation.

Hybrid extensions like View-opacity-Dependent 3D Gaussian Splatting (VoD-3DGS)~\cite{nowak2025vod} and Viewing Direction Gaussian Splatting (VDGS)~\cite{malarz2025gaussian} augment 3DGS with learnable matrices or classical MLPs to extend view-dependence to opacity. While effective, these classical augmentations remain expressivity-limited in the low-parameter regime required to preserve 3DGS's real-time performance.

Concurrently, Quantum Machine Learning (QML) has demonstrated unique advantages in high-dimensional function approximation. Prior quantum rendering efforts, Quantum Radiance Fields (QRF)~\cite{qrf} and QKAN-GS~\cite{qkan-gs}, focused on NeRF integration or 3DGS compression, leaving a critical gap: explicitly modeling view-dependent radiance via quantum-geometric encodings.

In this paper, we introduce \our{}, a novel quantum-classical hybrid that embeds Variational Quantum Circuits (VQCs) directly into the 3DGS pipeline to capture complex view-dependent dynamics. Our key strategy is a Bloch sphere encoding that maps 3D viewing directions onto qubit states, naturally representing view-dependent reflectance through single-qubit rotations. We propose dual VQC control strategies:
\begin{itemize}
\vspace{-0.3cm}
    \item \textbf{Pipeline I (Per-Gaussian):} A hypernetwork generates spatially-adaptive VQC parameters (rotation angles) for each Gaussian, maximizing local expressivity,
\vspace{-0.2cm}    
    \item \textbf{Pipeline II (Global):} A shared VQC with hash-conditioned global parameters scales to real-world scenes.
\end{itemize}
\vspace{-0.3cm}
This dual methodology allows us to explore the
trade-off between per-instance precision and global scalability.
Our key contributions are summarized as follows:
\vspace{-0.2cm}    
\begin{itemize}
\item \textbf{Bloch Sphere Directional Encoding:} We propose a novel embedding that maps viewing directions onto qubit states, yielding a physically meaningful representation of anisotropic reflections.
\vspace{-0.2cm}    
\item \textbf{Dual Quantum Control Mechanisms:} We develop two complementary schemes for modulating the parameters of VQCs through a classical hypernetwork or through global conditioning, enabling a controllable balance between local accuracy and large-scale generalization.
\vspace{-0.2cm}    
\item \textbf{Enhanced View-Dependent Rendering:} We show that our quantum–hybrid framework achieves improved expressivity and generalization, consistently surpassing conventional 3DGS and other classical baselines.
\vspace{-0.2cm}    
\end{itemize}

\section{Related Work}

\paragraph{Gaussian Splatting and Appearance Modeling}
3DGS~\cite{vanilla_3dgs} relies on low-order Spherical Harmonics (SH), which inherently act as a low-pass filter, failing to capture sharp specularities. While hybrid extensions like VDGS~\cite{malarz2025gaussian} introduce MLP modulation to mitigate this, they remain constrained by the limited expressivity of classical networks in low-parameter regimes, often resulting in blurred reflections for highly anisotropic materials.

\paragraph{Quantum Neural Rendering}
Despite rapid advances in Quantum Machine Learning (QML), its application to 3D neural rendering remains largely unexplored. The few pioneering works primarily target implicit scene representations.

Quantum Radiance Fields (QRF)~\cite{qrf} (2023) replaced NeRF~\cite{vanilla_nerf}, MLPs with Parameterized Quantum Circuits (PQCs) and quantum activation functions, enabling better capture of high-frequency details via higher-order derivatives. The authors further introduced quantum volume rendering via Grover's search to accelerate ray integration. However, QRF prioritized computational efficiency in numerical integration over modeling view-dependent geometric effects.

\paragraph{Quantum-Enhanced Gaussian Splatting}
Recent research has shifted alongside the broader community from implicit NeRFs to explicit 3DGS~\cite{vanilla_3dgs}. QKAN-GS~\cite{qkan-gs} introduced Quantum Kolmogorov-Arnold Networks (QKANs) with learnable QReLU activations at network edges, enabling compact representations of Gaussian attributes (opacity, covariance) with significantly fewer parameters than classical counterparts. Their approach emphasized storage efficiency rather than physically-motivated view-dependent appearance modeling.

\begin{figure*}[t]
\centering
\includegraphics[width=\linewidth]{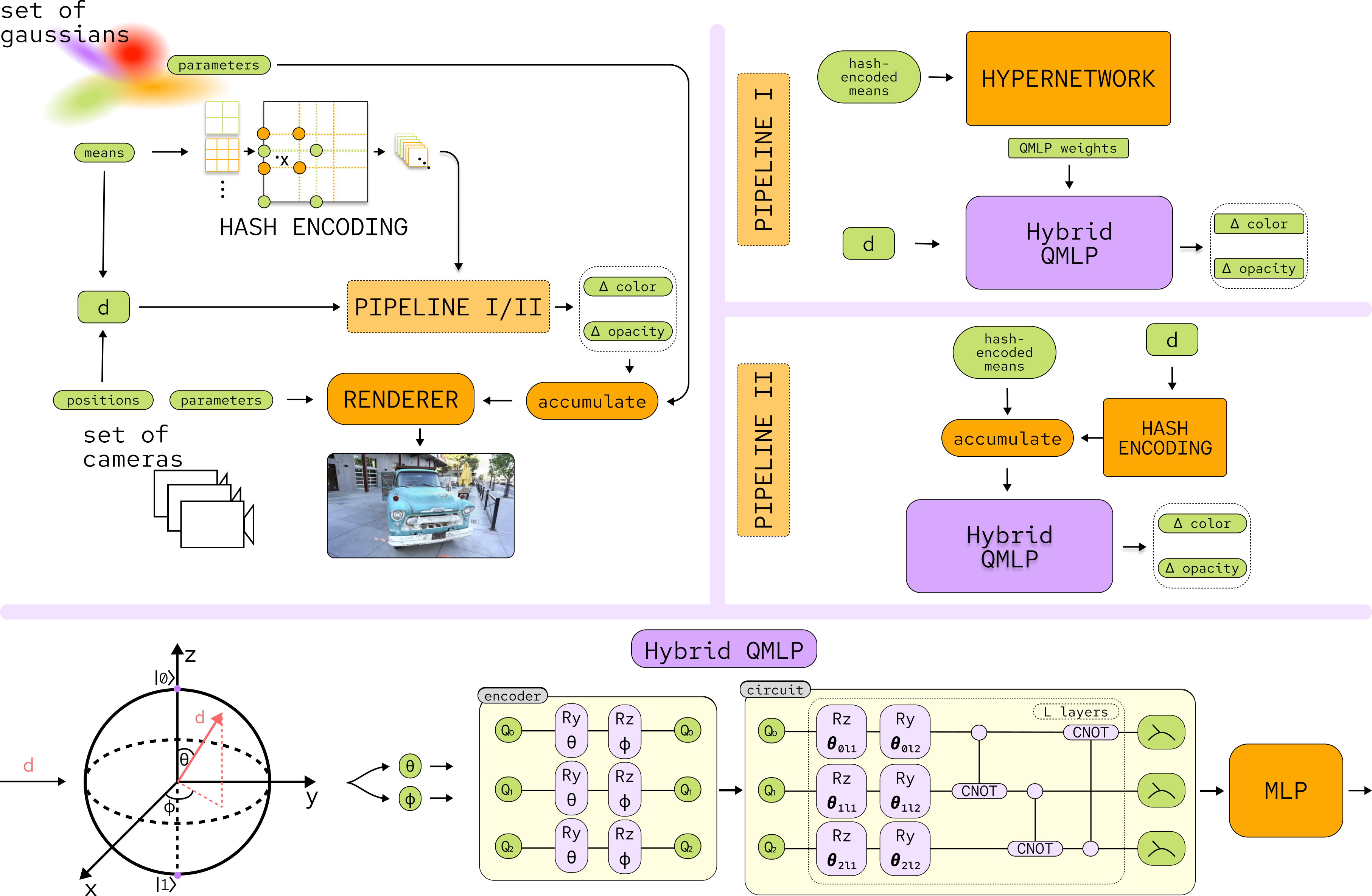}
\caption{
Our framework integrates quantum processing into 3D Gaussian Splatting for view-dependent color and opacity residuals. {\em Top}: Two interchangeable pipelines. Pipeline I (Hyper-Quantum) generates per-Gaussian VQC parameters via hypernetwork from spatial hash encoding. Pipeline II (Joint-Hash Global) feeds spatial and directional hash features to a shared quantum network. {\em Bottom}: Hybrid QMLP maps viewing directions to Bloch sphere via rotation gates ($R_y$, $R_z$), processes through VQC with circular entanglement, and decodes measurements via classical MLP to guide rendering.
}
\label{fig:method}
\end{figure*}

\paragraph{Differentiation}
While QKAN-GS demonstrates the utility of quantum networks for compression and attribute generation, it treats quantum layers primarily as parameter-efficient function approximators for static attributes. In contrast, \our{} addresses a fundamentally different challenge: modeling complex, view-dependent light-matter interactions. Unlike prior works that use quantum circuits merely as drop-in replacements for classical neurons, \our{} explicitly leverages quantum state space geometry. By mapping viewing directions directly to the Bloch sphere, it exploits qubits' natural rotational properties to model anisotropic effects and transparency, aspects unexplored by QRF or QKAN-GS.

\section{\our{}: Quantum Encoding Framework for Gaussian Splatting}
\label{methodology}
To overcome the limitations of classical neural networks in modeling complex, high-frequency view-dependent effects, we introduce \our{}, a hybrid framework that preserves the efficient 3D geometry of Gaussian Splatting while delegating light-matter interactions to a quantum neural network. Unlike traditional methods that treat viewing directions as discrete Euclidean coordinates, our approach embeds them directly into the Bloch sphere of a multi-qubit Hilbert space. This quantum-native representation captures rotational continuity, phase relationships, and interference effects to enable more expressive modeling of transparency, anisotropic reflections, and specular highlights, while remaining compatible with spherical harmonic color representations.

The \our{} architecture is illustrated schematically in Fig.~\ref{fig:method}. Below, we describe its methods and components in detail.

\subsection{Gaussian Splatting}

3D Gaussian Splatting (3DGS)~\cite{vanilla_3dgs} represents a scene as a collection of anisotropic 3D Gaussians
\begin{equation}
    \mathcal{G} = \{ (\mathcal{N}(\boldsymbol{\mu}, \boldsymbol{\Sigma}), \alpha, c) \},
\end{equation}
where each Gaussian is defined by its mean (center) $\boldsymbol{\mu}$, covariance 
$\boldsymbol{\Sigma} = \boldsymbol{R} \boldsymbol{S}\boldsymbol{S} \boldsymbol{R}^{\top}$ 
(with rotation $\boldsymbol{R}$ and scaling $\boldsymbol{S}$), opacity $\alpha$, and color $c$. To model view-dependent appearance, colors are represented as functions of viewing direction $\boldsymbol{d}$ via Spherical Harmonics (SH)~\cite{fridovich2022plenoxels,muller2022instant}:
\begin{equation}\label{eq:sh}
    c(\boldsymbol{d}) = \sum_{\ell,m} a_{\ell, m} \, Y_\ell^m(\boldsymbol{d}),
\end{equation}
enabling 3DGS to efficiently capture simple angular dependencies. Rendering projects 3D Gaussians onto the image plane and composites them via alpha blending~\cite{yifan2019differentiable}.
Recent improvements like VoD-3DGS~\cite{nowak2025vod} and VDGS~\cite{malarz2025gaussian} extend view-dependence to opacity $\alpha(\boldsymbol{d})$, using learnable symmetric matrix (VoD-3DGS) or MLP modulation (VDGS) for enhanced realism.

While this formulation provides real-time performance and robust geometric reconstruction, the spherical harmonic basis remains limited for modeling high-frequency, view-dependent, or specular effects. To overcome this, we introduce a quantum embedding mechanism that maps classical directional inputs into a continuous, rotation-aware~quantum~feature~space.

\subsection{Quantum Encoding of Viewing Direction}

Classical approaches typically represent viewing directions as points in 3D Cartesian coordinates. However, such representations fail to capture the continuous rotational topology and the underlying $\text{SO}(3)$ symmetry inherent to angular data defined on the unit sphere.

Quantum systems overcome this limitation by mapping a normalized viewing direction 
$\boldsymbol{d} = [d_x, d_y, d_z] \in \mathbb{S}^2 \subset \mathbb{R}^3$ 
onto the Bloch sphere, which is the geometric representation of pure single-qubit states in the complex Hilbert space $\mathcal{H}_2$. This mapping yields the Bloch coordinates:
\begin{equation}\label{eq:angles}
\begin{gathered}
    \theta = \arccos(d_z), \quad
    \phi = \arctan(d_y, d_x),
\end{gathered}
\end{equation}
with periodic normalization:
\begin{equation}\label{eq:angle_modulo}
    \phi = (\phi + 2\pi) \bmod 2\pi,
\end{equation}
which we call viewing angles. The corresponding qubit state admits the canonical Bloch-sphere parameterization:
\begin{equation}
    \ket{\psi} = \alpha\ket{0} 
    + \beta\ket{1} = \cos\!\left(\frac{\theta}{2}\right)\ket{0} 
    + e^{i\phi}\sin\!\left(\frac{\theta}{2}\right)\ket{1},
\end{equation}
where $|\alpha|^2$ and $|\beta|^2$ encode the probabilistic latitude of the state, 
while $e^{i\phi}$ determines its azimuthal phase (longitude). 
This formulation establishes a diffeomorphic mapping that faithfully preserves the manifold structure of $\mathbb{S}^2$.

For practical implementation, we extend this to a 3-qubit system 
($\mathcal{H}_2^{\otimes 3}$) using rotation gates:
\begin{equation}\label{eq:encoder}
    \ket{\psi_{\text{enc}}} 
    = \bigotimes_{j=0}^{2} R_z^{(j)}(\phi)\, R_y^{(j)}(\theta)\ket{0}^{\otimes 3},
\end{equation}
where $R^{(j)}_y(\theta) = \exp(-i\frac{\theta}{2}Y)$ encodes the polar (latitude) component 
and $R^{(j)}_z(\phi) = \exp(-i\frac{\phi}{2}Z)$ sets the azimuthal (longitude) phase for the $j$-th qubit ($Y$ and $Z$ denote Pauli matrices\footnote{Precisely, $Y=\begin{pmatrix} 0 & -i \\ i & 0 \end{pmatrix}$ and $Z=\begin{pmatrix} 1 & 0 \\ 0 & -1 \end{pmatrix}$.}). 
This encoding enables the network to interpret viewing directions as continuous rotations, 
providing a richer and more physically consistent basis for modeling light–surface interactions 
compared to discrete directional inputs used in standard NeRFs or VDGS frameworks.

\subsection{Variational Quantum Circuit}

The Bloch-encoded viewing direction $\ket{\psi_{\text{enc}}}$ is processed through a Variational Quantum Circuit (VQC) ansatz $U (\boldsymbol{\theta}; L)$ consisting of $L$ layers ($L=4$) parametrized by rotation angles $\boldsymbol{\theta}=(\theta_{j,\ell},\phi_{j,\ell})_{j=0,\ell=1}^{2,L}$, designed to introduce entanglement and trainable non-linearities. Each layer alternates between single-qubit rotations and multi-qubit entanglement to learn expressive, view-dependent transformations of the directional encoding.

Formally, the $\ell$-th layer $U^{(\ell)}(\boldsymbol{\theta}_\ell)$ of the ansatz, where $\boldsymbol{\theta}_\ell = (\theta_{j,\ell},\phi_{j,\ell})_{j=0}^{2}$ collects all its trainable parameters, applies the following sequence to the 3-qubit register:
\begin{enumerate}
    \item \textbf{Parameterized Rotations}: For the $j$-th qubit ($j=0,1,2$), local rotations
    \begin{equation}\label{eq:param_rot}
        R_j(\theta_{j,\ell},\phi_{j,\ell}) = R_z^{(j)}(\phi_{j,\ell}) R_y^{(j)}(\theta_{j,\ell})
    \end{equation}
    are applied. These layer-specific angles learn how different viewing directions transform the initial Bloch-sphere encoding.
    \item \textbf{Circular Entanglement}: A cyclic CNOT sequence
    \begin{equation}\label{eq:cnot}
        U_{\mathrm{ent}} = \mathrm{CNOT}_{0\to1} \; \mathrm{CNOT}_{1\to2} \; \mathrm{CNOT}_{2\to0}
    \end{equation}
    entangles all three qubits, creating correlations that model interdependencies between color channels (RGB) and opacity. This ring topology ensures view-dependent effects emerge from the joint quantum state.
\end{enumerate}
The full ansatz transformation is thus provided by the following formula: 
\begin{equation}\label{eq:ansatz}
    |\psi_{\mathrm{out}}\rangle=U(\boldsymbol{\theta}; L) |\psi_{\text{enc}}\rangle =  U^{(L)}(\boldsymbol{\theta}_L)\cdots U^{(1)}(\boldsymbol{\theta}_1) |\psi_{\text{enc}}\rangle,
\end{equation}
where
\begin{equation}\label{eq:ansatz_layer}
    U^{(\ell)}(\boldsymbol{\theta}_\ell)=U_{\mathrm{ent}} \bigotimes_{j=0}^2 R_j(\theta_{j,\ell},\theta_{j,\ell}).
\end{equation}
Computational basis measurements of the output state $|\psi_{\mathrm{out}}\rangle$ in the $Z$-basis (the standard $(|0\rangle,|1\rangle)$ basis of quantum hardware) yield expectation values $\langle Z_j \rangle$ for all qubits. 
The resulting 3D quantum feature vector $(\langle Z_0 \rangle, \langle Z_1 \rangle, \langle Z_2 \rangle) \in [-1,1]^3$ encodes nonlinear, entangled correlations from VQC processing and feeds into a lightweight classical MLP, yielding view-dependent refinements $\Delta c(\boldsymbol{d})$ and $\Delta\alpha(\boldsymbol{d})$ to the Gaussian's base color (SH) and opacity.

This yields a hybrid Quantum-MLP (QMLP) architecture, illustrated in the bottom part of Fig.~\ref{fig:method}, which combines quantum-native directional encoding with efficient 3DGS rasterization.

\subsection{Dual-Pipeline Framework: Local vs. Global Modeling}

\our{} employs two complementary pipelines balancing per-Gaussian expressivity with scene-scale scalability.

\paragraph{Pipeline I: Per-Gaussian Hyper-Quantum Modeling} A hypernetwork $\mathcal{H}$ generates unique VQC parameters (the rotation angles) $\boldsymbol{\theta}_\text{G}$
and MLP weights $\boldsymbol{W}^{\text{MLP}}_\text{G}$ for each Gaussian $\text{G}\in \mathcal{G}$ from its hashed position $\boldsymbol{\mu}_\text{G}$:
\begin{equation}\label{eq:hypernetwork}
    (\boldsymbol{\theta}_\text{G}, \boldsymbol{W}^{\text{MLP}}_\text{G}) = \mathcal{H}(\texttt{hash}(\boldsymbol{\mu}_\text{G})).
\end{equation}
These spatially-adaptive VQC parameters configure the per-Gaussian ansatz $U(\boldsymbol{\theta}_\text{G}; L)$, maximizing local precision for intricate optical effects like specular highlights on curved surfaces. This approach excels for synthetic scenes prioritizing high PSNR.

\begin{table}[t]
  \caption{Quantitative comparison of \our{} against state-of-the-art neural rendering methods on the NeRF Synthetic dataset. Using Pipeline I (Hyper-Quantum), \our{} achieves state-of-the-art performance with the highest PSNR (33.98) and SSIM (0.970) among all baselines, including standard 3DGS and VDGS. These results confirm the efficacy of Bloch sphere mapping and spatially-adaptive VQCs for modeling complex geometries and high-frequency specular effects.
  }
  \label{tab:nerf-synth}
  \begin{center}
    \begin{small}
      \begin{sc}
        \begin{tabular}{@{\;}l@{\;\;\;\;}c@{\;\;\;\;}c@{\;\;\;\;}c@{\;\;\;\;}c@{\;}}
          \toprule
          Method  & PSNR$\uparrow$        & SSIM$\uparrow$     & LPIPS~$\downarrow$     & FPS$\uparrow$ \\
          \midrule
          NeRF    & 31.01  & 0.947 & 0.081 & 0.023  \\
          VolSDF & 27.96  & 0.932 & 0.096 & ---  \\
          Ref-NeRF    & 31.29  & 0.947 & 0.058 & ---  \\
          ENVIDR    & 28.13  & 0.956 & 0.067 & ---  \\
          QRF     & 32.65  & \cellcolor{third}0.960 & \cellcolor{best} 0.029 & 47.26  \\
          \midrule
          GS      & \cellcolor{third}33.30  & \cellcolor{second}0.969 & \cellcolor{second}0.030 & 733.00  \\
          VDGS      & \cellcolor{second}33.37  & \cellcolor{second}0.969 & \cellcolor{third}0.032 & 284.29  \\
          \our{}   & \cellcolor{best}33.98  & \cellcolor{best}0.970 & \cellcolor{second}0.030 & 12.64  \\
          \bottomrule
        \end{tabular}
      \end{sc}
    \end{small}
  \end{center}
  \vskip -0.1in
\end{table}

\begin{table*}[t]
  \caption{Quantitative evaluation of \our{} against state-of-the-art neural rendering baselines on large-scale real-world datasets (Mip-NeRF 360, Tanks and Temples, Deep Blending). Using Pipeline II (Joint-Hash Global) with shared VQC conditioned on spatial+directional hash features, \our{} demonstrates superior generalization, leading in SSIM and LPIPS on Mip-NeRF 360 and top PSNR on Deep Blending.}
  \label{tab:big-scenes}
  \centering
  \begin{small}
  \begin{sc}
  \begin{tabular}{@{\;}l@{\;\;\,}c@{\;\;\,}c@{\;\;\,}c@{\;\;\,}c@{\;\;\,}c@{\;\;\,}c@{\;\;\,}c@{\;\;\,}c@{\;\;\,}c@{\;\;\,}c@{\;\;\,}c@{\;\;\,}c@{\;}}
    \toprule
    & \multicolumn{4}{c}{MiP-NeRF 360} 
    & \multicolumn{4}{c}{DeepBlending}
    & \multicolumn{4}{c}{Tanks and Temples} \\
    \cmidrule(lr){2-5}
    \cmidrule(lr){6-9}
    \cmidrule(lr){10-13}
    Method 
    & PSNR~$\uparrow$ & SSIM~$\uparrow$ & LPIPS~$\downarrow$ & FPS~$\uparrow$
    & PSNR~$\uparrow$ & SSIM~$\uparrow$ & LPIPS~$\downarrow$ & FPS~$\uparrow$
    & PSNR~$\uparrow$ & SSIM~$\uparrow$ & LPIPS~$\downarrow$ & FPS~$\uparrow$ \\
    \midrule
    Plenoxels
      & 23.08   & 0.626   & 0.719   & 6.79
      & 23.06   & 0.510   & 0.510   & 11.2
      & 21.08   & 0.379   & 0.795   & 13.0 \\
    INGP-Base
      & 25.30   & 0.671   & 0.371   & 11.7
      & 23.62   & 0.797   & 0.423   & 3.26
      & 21.72   & 0.723   & 0.330   & 17.1 \\
    INGP-Big
      & 25.59   & 0.699   & 0.331   & 9.43
      & 24.96   & 0.817   & 0.390   & 2.79
      & 21.92   & 0.745   & 0.305   & 14.4 \\
    MiPNeRF360
      & \cellcolor{best}27.69   & 0.792   & \cellcolor{third}0.237   & 0.06
      & 29.40   & 0.901   & \cellcolor{third}0.245   & 0.09
      & 22.22   & 0.759   & 0.257   & 0.14 \\
    QRF
      & ---   & ---   & ---   & ---
      & ---   & ---   & ---   & ---
      & \cellcolor{best}29.65   & 0.820   & \cellcolor{best}0.085   & 14.36 \\
    \midrule
    GS-30K
      & 27.21   & \cellcolor{best}0.815   & \cellcolor{best}0.214   & 134
      & \cellcolor{third}29.41   & \cellcolor{third}0.903   & \cellcolor{second}0.243   & 137
      & 23.14   & 0.841   & 0.183   & 154 \\
    VDGS
      & \cellcolor{second}27.64   & \cellcolor{second}0.813   & \cellcolor{second}0.220   & 41.35
      & \cellcolor{second}29.54   & \cellcolor{second}0.906   & \cellcolor{second}0.243   & 44.72
      & 24.02   & \cellcolor{third}0.851   & 0.176   & 28.53 \\
    QKAN-GS
      & ---   & ---   & ---   & ---
      & ---   & ---   & ---   & ---
      & \cellcolor{third}24.28   & \cellcolor{second}0.859   & \cellcolor{third}0.169   & --- \\
    \our{}
      & \cellcolor{third}27.27   & \cellcolor{third}0,793   & 0,244  & 10,78
      & \cellcolor{best}30.15   & \cellcolor{best}0.916   & \cellcolor{best}0.163   & 10.76
      & \cellcolor{second}24.70   & \cellcolor{best}0.888   & \cellcolor{second}0.118   & 16.15 \\
    \bottomrule
  \end{tabular}
  \end{sc}
  \end{small}
  \vskip -0.1in
\end{table*}






\begin{figure*}[t]
\centering
\includegraphics[width=\linewidth]{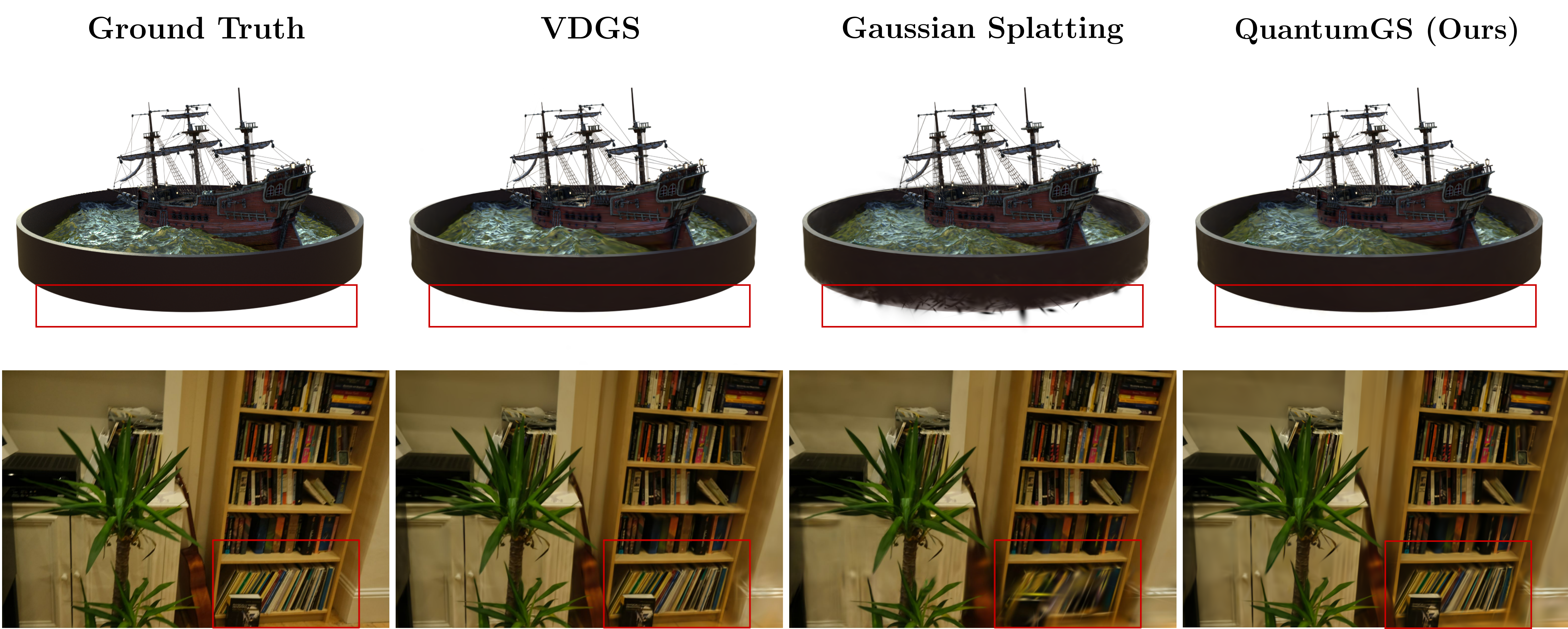}
\caption{
{\em Top}: The Ship scene (NeRF Synthetic) features thin rigging and liquid transparency. Standard 3DGS produces distracting ``floater'' artifacts beneath the model. \our{} generates a clean background comparable to ground truth. {\em Bottom}: In the Room scene (Mip-NeRF 360), standard 3DGS struggles with geometric consistency at the bookshelf base, creating jagged artifacts. \our{} eliminates these errors, preserving straight lines and structural coherence.
}
\label{fig:comparison_m_0}
\end{figure*}

\begin{figure*}[t]
\centering
\includegraphics[width=\linewidth]{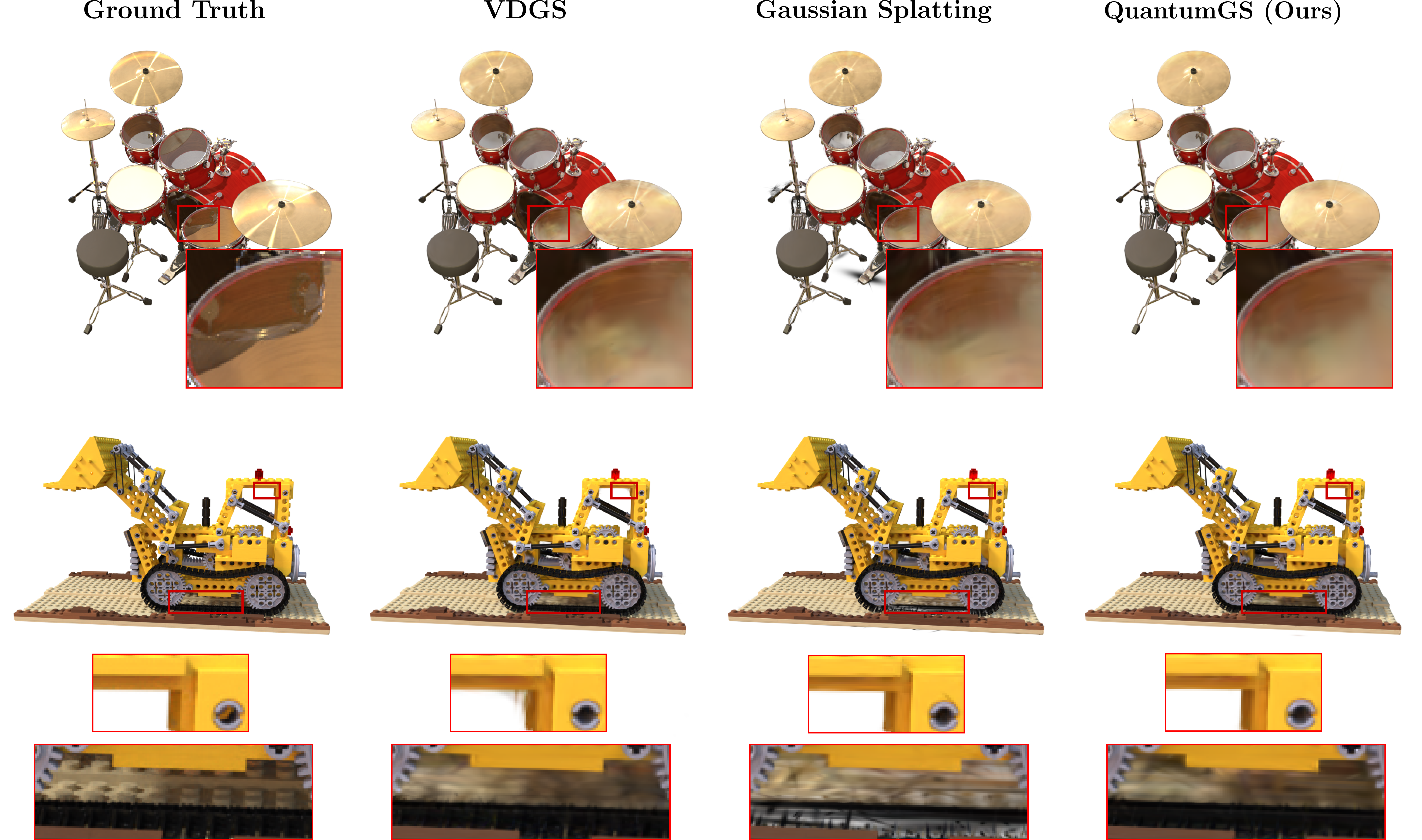}
\caption{Object-centric scenes. In the Drums scene, VDGS blurs the reflection on the drum surface, losing the geometric definition of the reflected drum. \our{} preserves the distinct shape of the reflection. In the LEGO scene, VDGS exhibits floater artifacts near the roof. Additionally, \our{} recovers occlusion shadows on the chassis, unlike standard 3DGS.}
\label{fig:comparison_zoom}
\end{figure*}

\begin{figure*}[t]
\centering
\includegraphics[width=\linewidth]{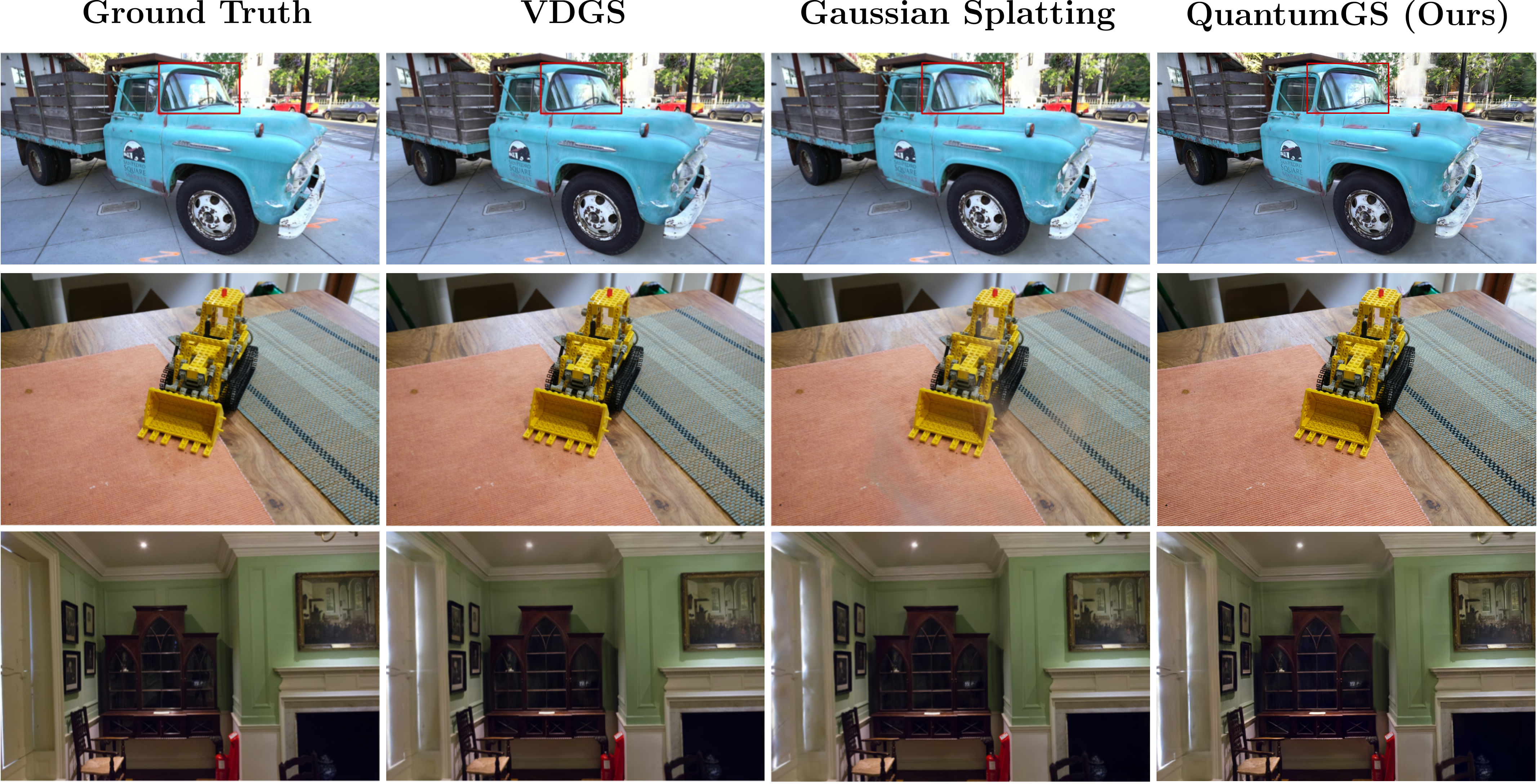}
\caption{Comparisons on real-world datasets. In the Truck scene (Tanks and Temples), standard 3DGS fails to capture high-frequency reflections on the windshield, resulting in a blurred appearance, whereas \our{} recovers sharp specular details. In the Kitchen scene (Mip-NeRF 360), standard 3DGS renders the LEGO truck with unnatural ``foggy'' or hazy appearance due to lighting ambiguity. \our{} resolves this issue, producing clear object boundaries. In the Dr. Johnson scene (Deep Blending), \our{} correctly handles high-dynamic-range light entering through the window, while standard 3DGS produces unnatural overexposure and artifacts on surrounding walls.}
\label{fig:comparison_m_1}
\end{figure*}

\paragraph{Pipeline II: Joint-Hash Global Modeling}

For large-scale real-world scenes, a shared VQC with global parameters $\boldsymbol{\theta}$ is conditioned on concatenated hash encodings of Gaussian positions $\texttt{hash}_\text{mean}(\boldsymbol{\mu})$ and viewing directions $\texttt{hash}_\text{vdir}(\boldsymbol{d})$. 
The quantum circuit serves as a scene-wide light field approximator with superior memory scaling while preserving quantum-enhanced view-dependent modeling. The output is then processed by a shared MLP with global weights $\boldsymbol{W}^{\text{MLP}}$.







\subsection{Differentiable Rendering and Optimization}

Final Gaussian attributes apply QMLP-generated view-dependent multiplicative updates to base SH color and opacity:
\begin{equation}\label{eq:updates}
    \mathbf{c}_{\text{final}}(\mathbf{d}) = \mathbf{c}(\mathbf{d}) \cdot \Delta c(\mathbf{d}), \;
    \alpha_{\text{final}}(\mathbf{d}) = \alpha(\mathbf{d}) \cdot \Delta\alpha(\mathbf{d}).
\end{equation}
These refined attributes feed into a differentiable Gaussian rasterizer for efficient 2D rendering from 3D primitives. The full pipeline optimizes end-to-end via a combined loss
\begin{equation}\label{eq:loss}
    \mathcal{L} = (1-\lambda)\mathcal{L}_1 + \lambda\mathcal{L}_{\text{D-SSIM}},
\end{equation}
which is scaled with a hyperparameter $\lambda\in (0,1)$. Training discovers optimal Bloch sphere rotations that eliminate view-dependent artifacts while faithfully reproducing complex scene radiance.

\section*{Experiments}

In this section, we present numerical experiments validating the effectiveness of our proposed \our{} framework. Evaluations were conducted on standard datasets, including: (a) the widely used NeRF Synthetic dataset for novel view synthesis (NVS) \cite{vanilla_nerf}, and (b) large-scale real-world scenes from Tanks and Temples \cite{knapitsch2017tanks}, Mip-NeRF 360 \cite{barron2022mip}, and Deep Blending \cite{DeepBlending2018}.

\paragraph{Quantitative Results} As shown in Table~\ref{tab:nerf-synth}, \our{} achieves state-of-the-art performance on NeRF Synthetic, attaining the highest PSNR and SSIM among all baselines, including 3DGS and its view-dependent extension VDGS. Per-scene results are detailed in Table~\ref{tab:nerfsynth-scenes}. Pipeline I consistently outperforms baselines across synthetic scenes, confirming the efficacy of spatially adaptive VQC parameters for complex geometries and high-frequency details.

For large-scale scenes using Pipeline II, results are summarized in Table~\ref{tab:big-scenes}. \our{} demonstrates superior generalization, surpassing 3DGS and VDGS on Deep Blending and Tanks and Temples in PSNR and SSIM. On Deep Blending, substantial gains highlight its robustness to complex lighting and unbounded scenes. On Mip-NeRF 360, it achieves top SSIM and LPIPS while matching PSNR, indicating that quantum-geometric encoding captures details elusive to spherical harmonics or MLP modulations.

\paragraph{Qualitative Results} Visual comparisons in Fig.~\ref{fig:teaser} and Figs.~\ref{fig:comparison_m_0}--\ref{fig:comparison_m_1} highlight our quantum encoding's advantages. In the Truck scene (Fig.~\ref{fig:teaser}), standard 3DGS's low-order spherical harmonics act as a low-pass filter, blurring the background poster visible through the windshield; \our{} recovers high-frequency view-dependent transparency and structural details.

In object-centric scenes (Fig.~\ref{fig:comparison_zoom}), our method excels in material definition. In the Drums scene, \our{} preserves distinct reflection geometry on the surface, unlike VDGS's blurred approximation. In the LEGO scene, it recovers chassis occlusion shadows and eliminates floating artifacts near the roof seen in baselines.

Geometric robustness appears in Fig.~\ref{fig:comparison_m_0}. In the Ship scene, \our{} eliminates distracting floaters beneath the model that plague 3DGS. In the Room scene, it maintains structural coherence at the bookshelf base, avoiding 3DGS's jagged artifacts.

For large-scale scenes (Fig.~\ref{fig:comparison_m_1}), \our{} better handles complex lighting. In the Kitchen scene, it removes unnatural haze around the LEGO object present in 3DGS. In Dr. Johnson's scene, it resolves high-dynamic-range window lighting, reducing overexposure on the surrounding geometry compared to 3DGS.

\paragraph{Performance and Real-Time Rendering}  Despite the computational overhead of quantum circuit simulation on classical hardware, \our{} maintains real-time rendering speeds. The efficient 3-qubit VQC design (utilizing lightweight matrix operations) successfully mitigates the costs typically associated with complex quantum mechanics simulation.

As shown in Tables~\ref{tab:nerf-synth} and~\ref{tab:big-scenes}, \our{} achieves frame rates ranging from 10 to over 16 FPS on high-resolution scenes. This demonstrates that the high expressivity of Bloch-sphere encoding does not preclude interactive visualization, enabling practical applications requiring both physical fidelity and responsiveness.

\paragraph{Implementation Details} The \our{} framework was implemented using PyTorch and 3DGS CUDA kernels for efficient rasterization. All experiments were conducted on a single NVIDIA GeForce RTX 4090 GPU. Optimization used the Adam optimizer for 30,000 iterations, following standard 3D Gaussian Splatting protocol.

\paragraph{Architectural Configurations} Both pipelines utilize multi-resolution hash encodings with $L=16$ levels, feature dimension $F=2$, and maximum hash table size $T=2^{19}$.
\begin{itemize}
\item Pipeline I (Synthetic): The hypernetwork is a classical MLP with two hidden layers of 64 units, residual connections, GeLU activation, and dropout. The final decoding MLP is lightweight: input layer of size 3, single hidden layer of 3 neurons, output layer of size $N_{out}$ (where $N_{out}=49$ for full SH+opacity modulation, or $N_{out}=1$ for opacity-only variants).
\item Pipeline II (Real-World): Projection networks following spatial and directional hashgrids are MLPs with 2 hidden layers of 64 units outputting dimension 3. The final global decoding MLP shares the lightweight structure of Pipeline I.
\end{itemize}

\paragraph{Hyperparameters and Training} Distinct training strategies were employed for each pipeline to address differences between object-centric synthetic data and large-scale environments.

For Pipeline I, learning rates for both the spatial $XYZ$ hash grid and hypernetwork were set to $5 \cdot 10^{-5}$. All other optimization parameters followed the original 3D Gaussian Splatting implementation.

For Pipeline II, adjustments ensured stability and memory efficiency: densification gradient threshold was $5\cdot10^{-4}$ to prevent excessive primitive growth. Learning rates were: spatial $XYZ$ hash grid and shared directional encoder at $1\cdot10^{-3}$; hybrid quantum network (VQC and decoding MLP) at $7.5\cdot10^{-3}$.

\begin{table}[t]
  \caption{Ablation study on NeRF Synthetic evaluating individual components of the \our{} framework. We compare Pipeline I (\our{} HYPER) and Pipeline II (\our{} GLOBAL MODEL) against variants with restricted quantum modulation (\our{} ONLY OPACITY or \our{} ONLY SH), and SH-free baseline (\our{} NO SH). Results show that joint view-dependent color and opacity residuals with per-Gaussian hypernetwork control achieve the highest rendering fidelity.
  }
  \label{tab:ablation}
  \begin{center}
    \begin{small}
      \begin{sc}
        \begin{tabular}{@{\;}l@{\;\;}c@{\;\;}c@{\;\;}c@{\;\;}c@{\;}}
          \toprule
          Method  & PSNR~$\uparrow$        & SSIM~$\uparrow$     & LPIPS~$\downarrow$\\
          \midrule
          \our{} no sh   & 33.06  & 0.967 & 0.036  \\
          \our{} only opacity & 33.45  & 0.968 & 0.034  \\
          \our{} global model  & 33.67  & 0.967 & 0.034  \\
          \our{} only sh & 33.87  & \textbf{0.970} & 0.031  \\
          \our{} hyper  & \textbf{33.98}  & \textbf{0.970} & \textbf{0.030}  \\
          \bottomrule
        \end{tabular}
      \end{sc}
    \end{small}
  \end{center}
  \vskip -0.1in
\end{table}

\paragraph{Ablation Study} Table~\ref{tab:ablation} evaluates the contribution of individual components and architectural choices within the \our{} framework on the NeRF Synthetic dataset.

First, we analyze the influence of base color representation. The \our{} NO SH variant sets spherical harmonics degree to 0 (leaving only RGB color vector per Gaussian), while our model predicts color and opacity changes. The performance drop compared to the full model indicates that retaining SH as base representation provides crucial initialization for view-dependent effects.

Next, we examine the scope of quantum modulation. In \our{} ONLY OPACITY, standard SH coefficients handle color while the quantum network predicts only opacity residuals. Conversely, \our{} ONLY SH modulates only color atop standard SH, leaving opacity unchanged. While single-attribute modulation improves over baselines, simultaneous color and opacity optimization yields the best results, confirming synergy between quantum-enhanced geometry and appearance.

Finally, we compare the two pipelines: \our{} GLOBAL MODEL (Pipeline II) vs. \our{} HYPER (Pipeline I). As shown in Table~\ref{tab:ablation}, the hypernetwork approach (HYPER) achieves the highest fidelity on object-centric synthetic scenes, validating per-instance quantum control for maximum local expressivity.

\section{Conclusions}

In this work, we introduced \our{}, a novel framework integrating explicit 3D rendering with Quantum Machine Learning. By replacing classical view-dependent functions with Variational Quantum Circuits and Bloch-sphere encoding, we demonstrate that quantum-geometric representations significantly outperform standard spherical harmonics for high-frequency optical effects.

Our dual-pipeline strategy balances local precision and global scalability: Pipeline I (Hyper-Quantum) achieves state-of-the-art fidelity on synthetic specular objects; Pipeline II (Joint-Hash Global) enables robust generalization in unbounded real-world scenes. Despite the classical simulation overhead, \our{} maintains interactive frame rates (10--16+ FPS), confirming the method's viability. This framework establishes a foundation for future neural rendering engines transitioning to real quantum hardware.

\paragraph{Limitations} The primary limitation is the cost of simulating quantum entanglement on classical GPUs, which currently prevents reaching the extreme rasterization speeds of vanilla 3DGS. While future NISQ hardware could bypass this bottleneck, the hypernetwork currently introduces higher optimization complexity than explicit coefficients. Balancing quantum expressivity with training efficiency remains a key focus for future work.

\nocite{langley00}

\bibliographystyle{icml2026}

\newpage
\appendix
\onecolumn
\section{Detailed Quantitative and Qualitative Results}
\label{appendix:detailed_results}

In this appendix, we provide a per-scene breakdown of our quantitative metrics on the NeRF Synthetic dataset and present additional visual comparisons to further substantiate the performance of \our{} across diverse scenarios.

\subsection{Per-Scene Quantitative Analysis}
Table~\ref{tab:nerfsynth-scenes} presents the detailed PSNR, SSIM, and LPIPS scores for each individual scene in the NeRF Synthetic dataset. \our{} demonstrates robust consistency, achieving the highest PSNR on \textbf{8 out of 8 scenes}, surpassing both the original 3DGS and the hybrid VDGS baseline. Notably, on challenging scenes with complex geometry and transparency such as \textit{Materials} and \textit{Hotdog}, our method shows significant improvements, validating the effectiveness of the proposed Pipeline I (Hyper-Quantum) in modeling high-frequency view-dependent effects.

\begin{table}[h!]
\caption{Per-scene results on NeRF Synthetic benchmark.}
\label{tab:nerfsynth-scenes}
\centering
\small
\begin{sc}
\setlength{\tabcolsep}{3pt}
\begin{tabular}{@{\;}l@{\;\;}c@{\;\;}c@{\;\;}c@{\;\;}c@{\;\;}c@{\;\;}c@{\;\;}c@{\;\;}c@{\;\;}c@{\;}}
\toprule
\multicolumn{10}{c}{\textbf{PSNR} $\uparrow$} \\
\midrule
Method & Chair & Drums & LEGO & Mic & Mat. & Ship & Hot. & Ficus & Avg. \\
\midrule
NeRF        & 33.00 & 25.01 & 32.54 & 32.91 & 29.62 & 28.65 & 36.18 & 30.13 & 31.01 \\
VolSDF      & 30.57 & 20.43 & 29.46 & 30.53 & 29.13 & 25.51 & 35.11 & 22.91 & 27.96 \\
Ref-NeRF    & 33.98 & \cellcolor{third}25.43 & 35.10 & 33.65 & 27.10 & 29.24 & 37.04 & 28.74 & 31.29 \\
ENVIDR      & 31.22 & 22.99 & 29.55 & 32.17 & 29.52 & 21.57 & 31.44 & 26.60 & 28.13 \\
\midrule
GS          & \cellcolor{third}35.82 & \cellcolor{second}26.17 & \cellcolor{second}35.69 & \cellcolor{second}35.34 & \cellcolor{third}30.00 & \cellcolor{third}30.87 & \cellcolor{third}37.67 & \cellcolor{third}34.83 & \cellcolor{third}33.30 \\
VDGS        & \cellcolor{second}35.97 & \cellcolor{second}26.17 & \cellcolor{third}35.40 & \cellcolor{third}34.76 & \cellcolor{second}30.67 & \cellcolor{second}30.94 & \cellcolor{second}38.04 & \cellcolor{second}34.98 & \cellcolor{second}33.37 \\
\our{}      & \cellcolor{best}35.86 & \cellcolor{best}26.33 & \cellcolor{best}36.30 & \cellcolor{best}36.64 & \cellcolor{best}30.90 & \cellcolor{best}31.86 & \cellcolor{best}38.17 & \cellcolor{best}35.81 & \cellcolor{best}33.98 \\
\bottomrule

\\[-0.4em]
\multicolumn{10}{c}{\textbf{SSIM} $\uparrow$} \\
\midrule
Method & Chair & Drums & LEGO & Mic & Mat. & Ship & Hot. & Ficus & Avg. \\
\midrule
NeRF        & 0.967 & 0.925 & 0.961 & 0.980 & 0.949 & 0.856 & 0.974 & \cellcolor{third}0.964 & 0.947 \\
VolSDF      & 0.949 & 0.893 & 0.951 & 0.969 & 0.954 & 0.842 & 0.972 & 0.929 & 0.932 \\
Ref-NeRF    & 0.974 & 0.929 & \cellcolor{third}0.975 & 0.983 & 0.921 & \cellcolor{third}0.864 & \cellcolor{third}0.979 & 0.954 & 0.947 \\
ENVIDR      & \cellcolor{third}0.976 & 0.930 & 0.961 & 0.984 & \cellcolor{best}0.968 & 0.855 & 0.963 & \cellcolor{second}0.987 &\cellcolor{third} 0.956 \\
\midrule
GS          & \cellcolor{second}0.987 & \cellcolor{second}0.954 & \cellcolor{best}0.983 & \cellcolor{second}0.991 & 0.960 & \cellcolor{best}0.907 & \cellcolor{second}0.985 & \cellcolor{second}0.987 & \cellcolor{second}0.969 \\
VDGS        & \cellcolor{second}0.987 & \cellcolor{third}0.950 & \cellcolor{second}0.981 & \cellcolor{third}0.990 & \cellcolor{second}0.965 & \cellcolor{second}0.903 & \cellcolor{second}0.985 & \cellcolor{second}0.987 & \cellcolor{second}0.969 \\
\our{}      & \cellcolor{best}0.988 & \cellcolor{best}0.955 & \cellcolor{best}0.983 & \cellcolor{best}0.992 & \cellcolor{third}0.963 & \cellcolor{best}0.907 & \cellcolor{best}0.986 & \cellcolor{best}0.988 & \cellcolor{best}0.970 \\
\bottomrule

\\[-0.4em]
\multicolumn{10}{c}{\textbf{LPIPS} $\downarrow$} \\
\midrule
Method & Chair & Drums & LEGO & Mic & Mat. & Ship & Hot. & Ficus & Avg. \\
\midrule
NeRF        & 0.046 & 0.091 & 0.050 & 0.028 & 0.063 & 0.206 & 0.121 & 0.044 & 0.081 \\
VolSDF      & 0.056 & 0.119 & 0.054 & 0.191 & 0.048 & 0.191 & 0.043 & 0.068 & 0.096 \\
Ref-NeRF    & 0.029 & 0.073 & \cellcolor{third}0.025 & \cellcolor{third}0.018 & 0.078 & 0.158 & \cellcolor{third}0.028 & 0.056 & \cellcolor{third}0.058 \\
ENVIDR      & 0.031 & 0.080 & 0.054 & 0.021 & 0.045 & 0.228 & 0.072 & \cellcolor{best}0.010 & 0.067 \\
\midrule
GS          & \cellcolor{second}0.012 & \cellcolor{best}0.037 & \cellcolor{best}0.016 & \cellcolor{best}0.006 & \cellcolor{second}0.034 & \cellcolor{best}0.106 & \cellcolor{best}0.020 & \cellcolor{third}0.012 & \cellcolor{best}0.030 \\
VDGS        & \cellcolor{third}0.013 & \cellcolor{third}0.042 & \cellcolor{second}0.018 & \cellcolor{second}0.008 & \cellcolor{best}0.032 & \cellcolor{third}0.113 & \cellcolor{second}0.022 & \cellcolor{third}0.012 & \cellcolor{second}0.032 \\
\our{}      & \cellcolor{best}0.010 & \cellcolor{best}0.037 & \cellcolor{best}0.016 & \cellcolor{best}0.006 & \cellcolor{third}0.035 & \cellcolor{second}0.107 & \cellcolor{best}0.020 & \cellcolor{second}0.011 & \cellcolor{best}0.030 \\
\bottomrule
\end{tabular}
\end{sc}
\vskip -0.1in
\end{table}

\subsection{Additional Qualitative Comparisons}
\begin{figure*}[h!]
\centering
\includegraphics[width=\linewidth]{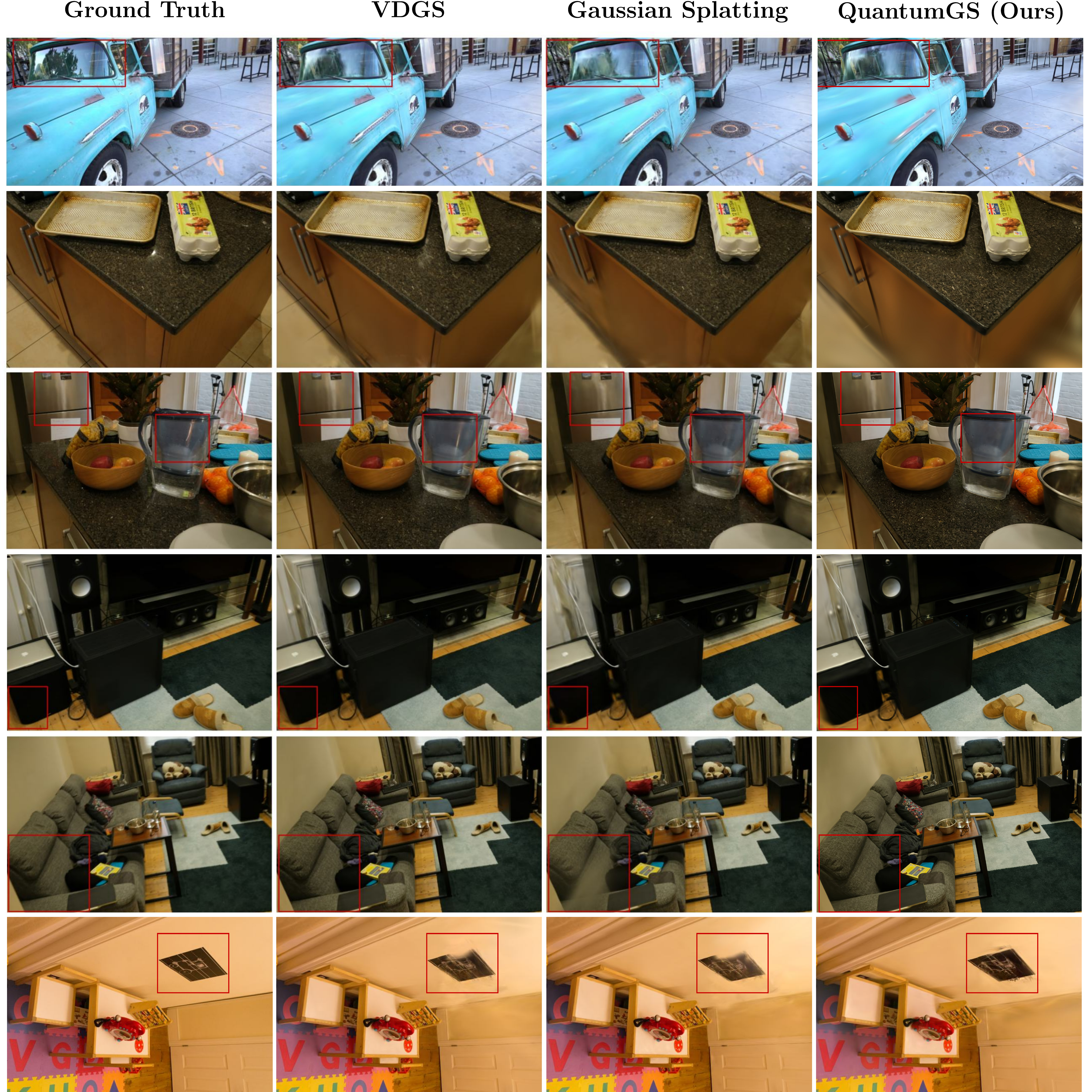}
\caption{
\textbf{Additional qualitative comparisons on Real-World scenes.} From top to bottom: \textit{Truck} (Tanks\&Temples), \textit{Counter} (Mip-NeRF 360), \textit{Room} (Mip-NeRF 360), and \textit{Playroom} (Deep Blending). Standard 3DGS struggles with reflections (Truck, Counter) and produces floaters (Room). \our{} consistently recovers fine details, such as the chalk drawings in the \textit{Playroom} scene, surpassing both baselines.
}
\label{fig:app_comparison_real}
\end{figure*}

\begin{figure*}[h!]
\centering
\includegraphics[width=\linewidth]{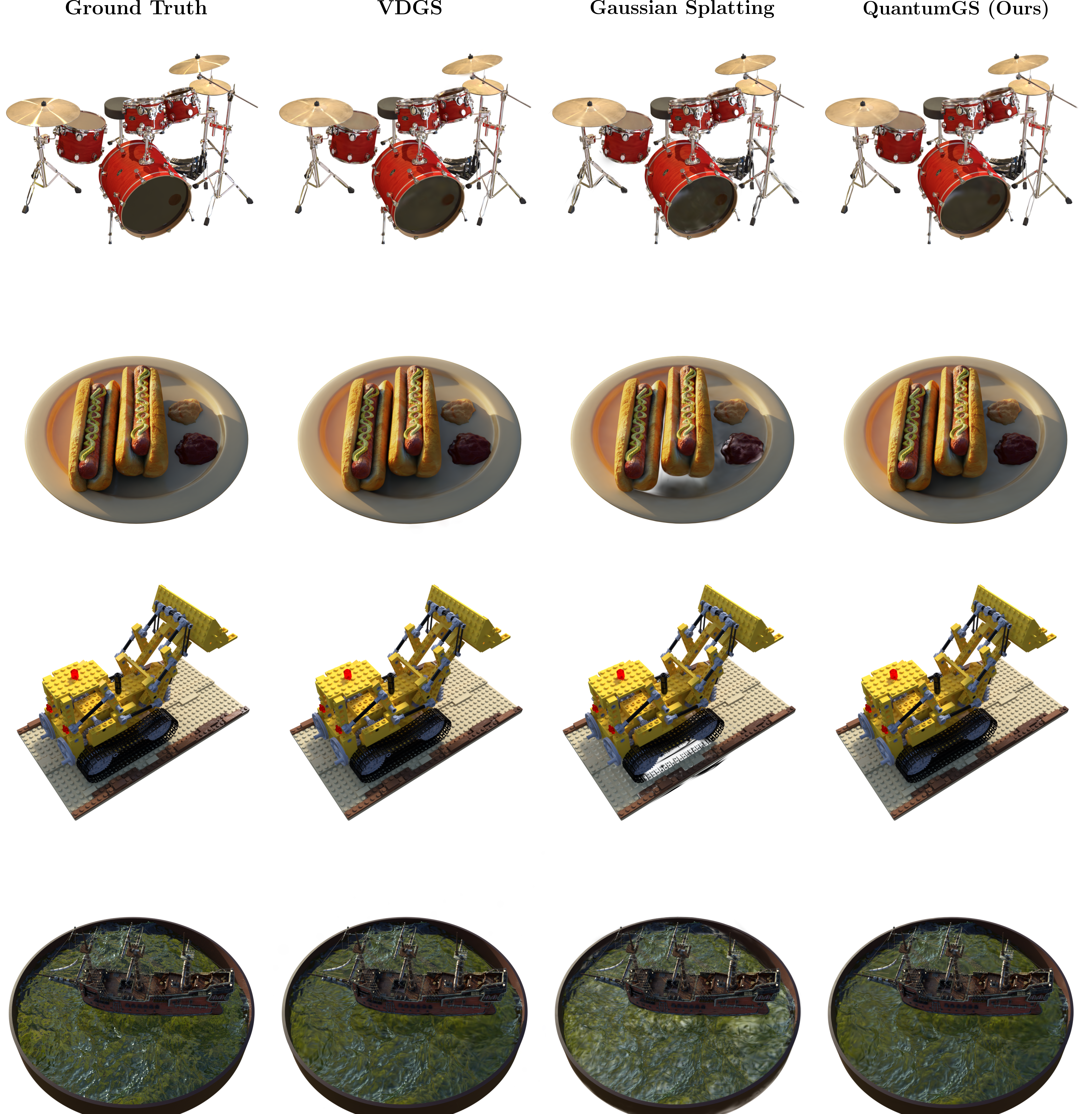}
\caption{
\textbf{Additional qualitative comparisons on Synthetic scenes.} From top to bottom: \textit{Drums}, \textit{Hotdog}, \textit{Lego}, and \textit{Ship}. Standard 3DGS exhibits visible artifacts, including blurred specular highlights on the drums, white halos around the hotdog and lego bulldozer, and spiky geometry near the water surface in the ship scene. \our{} matches or exceeds the quality of VDGS, effectively eliminating these high-frequency artifacts.
}
\label{fig:app_comparison_synth}
\end{figure*}

We provide extensive visual comparisons in Fig.~\ref{fig:app_comparison_real} and Fig.~\ref{fig:app_comparison_synth}.

\textbf{Real-World Scenes (Fig.~\ref{fig:app_comparison_real}).} On the \textit{Truck} scene, standard 3DGS fails to resolve the reflections on the windshield, resulting in a blurred appearance, while \our{} recovers specular details. In the \textit{Counter} scene, our method better preserves the sharpness of reflections compared to the baseline. In the \textit{Room} scene, standard 3DGS exhibits blurring artifacts caused by geometric "floaters," which are mitigated in our approach. Notably, in the \textit{Playroom} scene (Deep Blending), \our{} outperforms both 3DGS and VDGS in recovering fine details, such as the drawings on the chalkboard, which appear washed out in competing methods.

\textbf{Synthetic Scenes (Fig.~\ref{fig:app_comparison_synth}).} On the NeRF Synthetic dataset, both \our{} and VDGS significantly outperform standard 3DGS. Standard 3DGS struggles with glossy surfaces, producing artifacts on the \textit{Drums}. It also exhibits characteristic "white halo" artifacts near object boundaries in the \textit{Hotdog} and \textit{Lego} scenes. Furthermore, in the \textit{Ship} scene, standard 3DGS fails to model water reflections accurately, generating spiky Gaussian artifacts and white background bleeding. \our{} effectively eliminates these artifacts, producing clean and sharp renders.

\end{document}